\documentclass{article}
\usepackage{spconf,amsmath,graphicx}
\graphicspath{{figure/}}
\usepackage{amssymb}
\usepackage{url,hyperref,lineno,microtype,subcaption}
\usepackage{cite}

\usepackage{enumitem}
\setlist{nosep, leftmargin=14pt}

\usepackage{mwe} % to get dummy images

\title{Accelerating Diffusion Models via Pre-segmentation Diffusion Sampling for Medical Image Segmentation}
\name{Xutao Guo\textsuperscript{1,3}, Yanwu Yang\textsuperscript{1,3}, Chenfei Ye\textsuperscript{2}, Shang Lu \textsuperscript{1}, Yang Xiang\textsuperscript{3,*}, Ting Ma\textsuperscript{1,3,*} \thanks{\textsuperscript{*}Corresponding author: tma@hit.edu.cn, xiangy@pcl.ac.cn}}
\address{School of Electronics and Information Engineering, Harbin Institute of Technology at Shenzhen,
	\\ Shenzhen, China \\
		International Research Institute for Artifcial Intelligence, Harbin Institute of
		Technology at Shenzhen, \\ Shenzhen, China \\Peng Cheng Laboratory, Shenzhen, China}
\begin{document}
%\ninept
%
\maketitle
\begin{abstract}

Based on the Denoising Diffusion Probabilistic Model (DDPM), medical image segmentation can be described as a conditional image generation task, which allows to compute pixel-wise uncertainty maps of the segmentation and allows an implicit ensemble of segmentations to boost the segmentation performance. However, DDPM requires many iterative denoising steps to generate segmentations from Gaussian noise, resulting in extremely inefficient inference. To mitigate the issue, we propose a principled acceleration strategy, called pre-segmentation diffusion sampling DDPM (PD-DDPM), which is specially used for medical image segmentation. The key idea is to obtain pre-segmentation results based on a separately trained segmentation network, and construct noise predictions (non-Gaussian distribution) according to the forward diffusion rule. We can then start with noisy predictions and use fewer reverse steps to generate segmentation results. Experiments show that PD-DDPM yields better segmentation results over representative baseline methods even if the number of reverse steps is significantly reduced. Moreover, PD-DDPM is orthogonal to existing advanced segmentation models, which can be combined to further improve the segmentation performance.

\end{abstract}
\section{Introduction}
\label{sec:intro}

Denoising Diffusion Probabilistic Models (DDPM) \cite{10061924,10061925} form a category of deep generative models which has recently become one of the hottest topic in computer vision, due to their promising results in both unconditional and conditional generation tasks \cite{10061925,10061926,10061927}. Based on the Denoising Diffusion Probabilistic Model (DDPM), medical image segmentation can be described as a conditional image generation task, which allows to compute pixel-wise uncertainty maps of the segmentation and allows an implicit ensemble of segmentations to boost the segmentation performance \cite{10061928}. Especially in medical applications where subsequent diagnosis or treatment relies on segmentation, algorithms that provide only the most likely hypotheses can lead to misdiagnosis and suboptimal treatment. If multiple consistent hypotheses are provided, they can be used to suggest further diagnostic tests to resolve ambiguity, or experts with access to additional information can select appropriate hypotheses for subsequent steps \cite{10061951,10061928}.

The Denoising Diffusion Probabilistic Model (DDPM) consists of two Markov chains. In the forward diffusion process, the clean images are gradually disturbed by Gaussian noise until they are approximated to Gaussian distribution \cite{10061924}. In the process of inverse diffusion, from the sampled Gaussian noise, the trained denoising deep neural networks is used to iteratively denoise the data to obtain the clean images. Therefore, synthesizing samples from DDPM is achieved by iteratively denoising the sampled Gaussian noise. For the medical image segmentation, we can train the DDPM on the ground truth segmentation, and use the image as a prior during training and in every step during the sampling process \cite{10061928}. However, a major problem with vanilla DDPM is the inefficiency of inference. Because obtaining clean segmentations from DDPM typically requires hundreds or even thousands of denoising steps, each of which involves forward prediction by the denoising neural network.

\begin{figure*}[!t]
	\centering 
	%\vspace{-1cm}  %调整图片与上文的垂直距离
	%\setlength{\abovecaptionskip}{0.3cm}   %调整图片标题与图距离
	%\setlength{\belowcaptionskip}{0.0cm}   %调整图片标题与下文距离
	%\centerline{\includegraphics[width=0.72\textwidth]{fig03.eps}}
	%\centerline{\includegraphics[width=\columnwidth]{02.png}}
	%\centering%trim={<left> <lower> <right> <upper>}
	\includegraphics[scale=0.61,trim=0.15cm 6.9cm 0cm 4.6cm,clip]{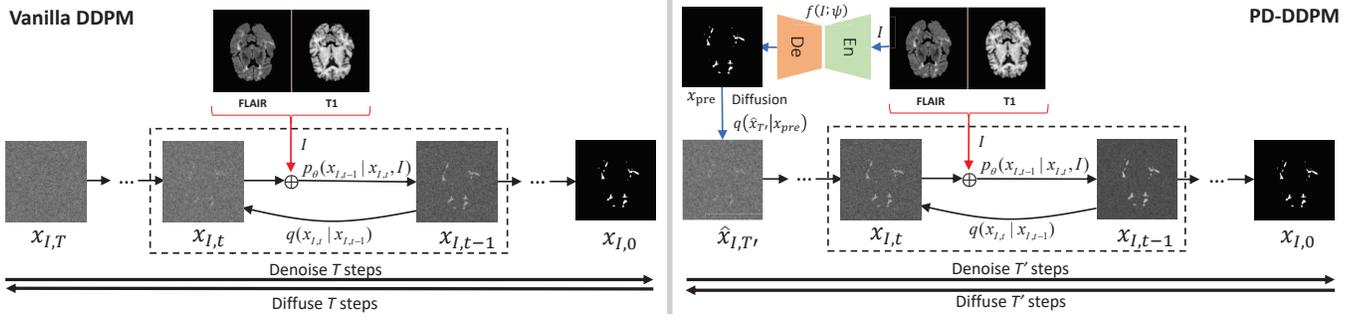}\\
	\caption{Compare PD-DDPM with the vanilla DDPM. The training and sampling procedure of the method. In every step $t$, the conditional information is induced by concatenating the medical images $I$ to the noisy segmentation mask $x_{I,t}$.}
	\label{fig001}
\end{figure*}

To accelerate DDPMs, a few methods have been proposed. Song et al. \cite{10251526} attempted to reduce the number of diffusion steps by using non-Markovian reverse processes. Unlike DDPM, which defines noise scale as a constant, San-Roman et al. \cite{10251543} present a adaptive noise scheduling to estimate the noise parameters given the current input at inference time, which requiring less steps. In \cite{10062032}, the authors propose to distill the full sampling process into a faster sampler that requires only half as many steps. Some methods \cite{10062033,10251600} shifted the diffusion process to the latent space using pre-trained autoencoders. However, the above methods cannot achieve significant acceleration without sacrificing the quality of power generation. Some other methods \cite{10062040,10062041} improve sampling efficiency by truncating the forward and reverse diffusion processes, boosting the performance at the same time. But this method needs to combine GAN\cite{10251613} or VAE\cite{10251612} models that are difficult to train. And it will break the characteristic that the image dimension remains unchanged during diffusion and sampling in the vanilla DDPM. To sum up, all of the above methods do not implement accelerated sampling specifically for segmentation tasks. 

To mitigate the issue, we propose a principled acceleration strategy, called pre-segmentation diffusion sampling DDPM (PD-DDPM), which is specially used for medical image segmentation. The key idea is to obtain pre-segmentation results based on a separately trained segmentation network, and construct noise predictions (non-Gaussian distribution) according to the forward diffusion rule. We can then start with noisy predictions and use fewer reverse steps to generate segmentation results. PD-DDPM not only improves the efficiency of vanilla DDPM without breaking any assumptions, but as an additional benefit, improves the segmentation performance of vanilla DDPM. Experiments show that PD-DDPM yields better segmentation results over representative baseline methods even if the number of reverse steps is significantly reduced. Moreover, PD-DDPM is orthogonal to existing advanced segmentation models, which can be combined to further improve the segmentation performance. Since both pre-segmentation and diffusion operations can be easily implemented in one step, the pre-trained segmentation network brings only minor computational overhead to PD-DDPM.

\section{METHOD}
\label{sec:format}

\subsection{Background on DDPMs}
\label{ssec:subhead}

In DDPM \cite{10061925}, the forward diffusion process is a first-order Markov chain that perturbs the data distribution $q(x^0)$ by gradually adding Gaussian noise, with variance $\beta_t\in(0,1)$ at time $t$, until the data distribution converges to a standard Gaussian distribution. The form of the forward process can be summarized as follows:
\begin{equation}
	\begin{split}
	q(x^{1:T}|x^0) &=\prod_{t=1}^{T}q(x^t|x^{t-1}),\\ 
	q(x^t|x^{t-1}) &= \mathcal{N}(x^t;\sqrt{1-\beta_t}x^{t-1},\beta_t\emph{\textbf{I}})
	\label{eq0001}
	\end{split}
\end{equation}
where $x^{1:T}$ denotes the set of variables $x^1, x^2, ..., x^T$. $T = 1000\backsim4000$ is a typical choice for most works. With the limit of small diffusion rate (i.e., $\beta_t$ is kept sufficiently small), the reverse distribution $q(x^{t-1}|x^t)$ also follows a Gaussian distribution. And thus the reverse process can be approximated using a neural network parameterized Gaussian distribution $p_\theta$, starting at $p(x^T) = \mathcal{N}(x^T;0,I)$:
\begin{equation}
	\begin{split}
		p_\theta(x^{0:(T-1)}|x^T) &=\prod_{t=1}^{T}p_\theta(x^{t-1}|x^t),\\ 
		p_\theta(x^{t-1}|x^t) &= \mathcal{N}(x^{t-1};\mu_\theta(x^t,t),\sigma^2_t\emph{\textbf{I}})
		\label{eq01}
	\end{split}
\end{equation}
The neural network is trained to simulate the reverse process of the diffusion process defined in Equation \ref{eq0001}. To generate an image from the reverse process, we first sample $x^T$ from the underlying data distribution by sampling a latent (of the same size as the training data point $x^0$) from $p(x^T)$ (chosen to be an isotropic Gaussian distribution), and then sequentially draws sample $x^{t-1}$ from $p_\theta(x^{t-1}|x^t)$ for $t = T,T-1,...,1$ until we get a new data $x^0$.The generation process of a DDPM is extremely slow as it need to sample from the transition distribution $p_\theta(x^{t-1}|x^t)$ iteratively, which involves a lot of evaluations of the output of the neural network.

\subsection{Pre-segmentation Diffusion Sampling Denoising Diffusion Probabilistic Model}
\label{ssec:subhead}

One natural question raised from the vanilla DDPM could be: can we cut the reverse process to $T'$ ($<T$) steps? If we can get the noise samples $x^{T'}$ in advance, it can be achieved. To generate image-specific segmentations, we train DDPM on the ground truth and use images as priors during training and sampling. To generate image-specific segmentation results, we train DDPM on the ground truth segmentation and use medical images as conditional information during training and sampling process. Therefore, we can train a separate segmentation network $f_\psi$ using medical image conditional information.
\begin{equation}
	\begin{split}
		x_{pre} = f(I;\psi)
		\label{eq01}
	\end{split}
\end{equation}

In the sampling process, we can first obtain the pre-segmentation result through the pre-segmentation network $f_\psi$ as shown in Figure \ref{fig001}. Then, according to Formula \ref{eq0001}, the pre-segmentation result is diffused to $T'$ step to obtain the approximate sample $\hat{x}^{T'}$ of $x^{T'}$. In this case the reverse process can be newly defined as:
\begin{equation}
	\begin{split}
		p_\theta(x^{0:(T'-1)}|\hat{x}^{T'}) &=\prod_{t=1}^{T'-1}p_\theta(x^{t-1}|x^t)p_\theta(x^{T'-1}|\hat{x}^{T'}),\\ 
		p_\theta(x^{t-1}|x^t) &= \mathcal{N}(x^{t-1};\mu_\theta(x^t,t),\sigma^2_t\emph{\textbf{I}})
		\label{eq0002}
	\end{split}
\end{equation}
Thus, we can use the denoising neural network to denoise the non Gaussian distribution $\hat{x}^{T'}$ to a clean segmentation $x^0$ in fewer steps than the vanilla sampling process according to Formula \ref{eq0002}. Because the pre-segmentation result is not the real ground truth segmentation, there will be some errors between $\hat{x}^{T'}$ and $x^{T'}$. However, we empirically found through experiments that PD-DDPM boosts the segmentation accuracy of vanilla DDPM. We name the pre-segmentation diffusion sampling DDPM as PD-DDPM. It should be emphasized that PD-DDPM does not break the Gaussian assumption of vanilla DDPM’ denoising process. In this way, the number of denoising steps required is greatly reduced, thus effectively alleviating the pain point of high inference cost of DDPM.

\section{EXPERIMENTS AND RESULTS}
\label{sec:pagestyle}

\subsection{Datasets}
\label{ssec:subhead}

We evaluate our method on the WMH dataset provided by the White Matter Hyperintensities segmentation challenge in MICCAI 2017 \cite{10071213}. It consists of 60 cases of brain MRI images (3D T1-weighted image and 2D multi-slice FLAIR image) with manual annotations of white matter hyperintensity (binary masks) from three different institutes/scanners. And the manual reference standard is defined on the FLAIR image. So a 2D multi-slice version of the T1 image was generated by re-sampling the 3D T1-weighted image to match with the FLAIR. In this paper, all cases are randomly assigned into five folds. Then we randomly assign these five-folds into a training set (3-fold), a validation set (1-fold), and a test set (1-fold).

\subsection{Implementation Details}
\label{ssec:subhead}

In this paper, all the networks train using Pytorch using NVIDIA TESLA V-100 (Pascal) GPUs with 32 GB memory. We optimized all configurations with the Adam optimizer with the learning rate 1e-4 and the weight decay 1e-5. The batch size is set to 12. For the WMH, images and annotation labels were randomly cropped to 128 × 192 patches. We choose a cosine noise schedule for $T=1000$ steps. The number of channels in the first layer of denoising network is chosen as $128$, and we use one attention head at resolution $16$. In PD-DDPM, we choose the AttUnet\cite{10081650} as the pre-segmentation model.

\begin{figure}[!t]
	\centering 
	%\vspace{-1cm}  %调整图片与上文的垂直距离
	%\setlength{\abovecaptionskip}{0.3cm}   %调整图片标题与图距离
	%\setlength{\belowcaptionskip}{0.0cm}   %调整图片标题与下文距离
	%\centerline{\includegraphics[width=0.72\textwidth]{fig03.eps}}
	%\centerline{\includegraphics[width=\columnwidth]{02.png}}
	%\centering%trim={<left> <lower> <right> <upper>}
	\includegraphics[scale=0.66,trim=1.66cm 1.1cm 0cm 0.9cm,clip]{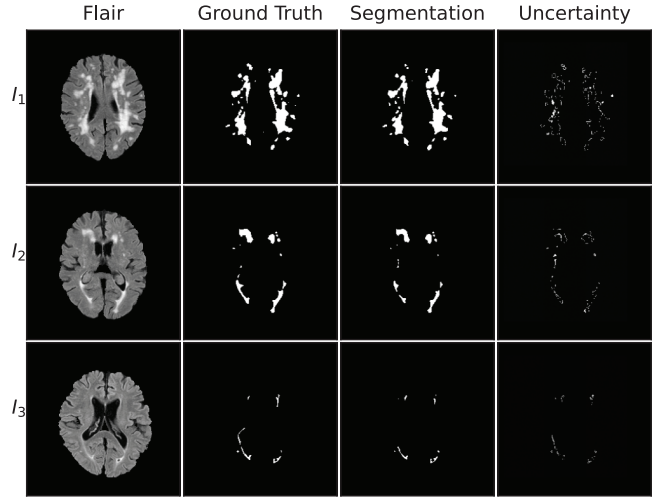}\\
	\caption{Visualization of segmentation and uncertainty maps.}
	\label{fig002}
\end{figure}

\begin{table}
	%\footnotesize
	\centering
	\caption{Segmentation scores with different methods.}
	\label{table}
	\setlength{\tabcolsep}{4.6pt}
	\begin{tabular}{c|cccc}
		\hline 			
		Methods       & Dice& HD95& Jaccard& F1\\
		\hline
		%\hline
		U-Net\cite{10071153}         &  0.787&  3.935&  0.656&  0.738\\
		%\hline
		AttUnet\cite{10081650}       &  0.799&  4.190&  0.673&  0.753\\
		%\hline
		U-Net++\cite{10081651}       &  0.798&  3.915&  0.672&  0.751\\
		\hline
		Bayesian U-Net\cite{10081652}		&  0.798&  3.803&  0.672&  0.777\\
		Probabilistic U-Net\cite{10241541}		    &  0.792&  3.777&  0.663&  0.763\\
		\hline
		Vanilla DDPM\cite{10061928}         &  0.796&  4.179&  0.669&  0.779\\
		\hline
		TDPM\cite{10062040}                 &  0.801&  3.608&  0.676&  0.776\\
		ES-DDPM\cite{10062041}              &  0.803&  3.523&  0.678&  0.777\\	
		%\hline
		\hline
		PD-DDPM                             &  \bf{0.812}&  \bf{3.494}&  \bf{0.689}&  \bf{0.800}\\
		\hline
	\end{tabular}
	\label{tab01}
\end{table}

\subsection{Comparison of Segmentation Performance}
\label{ssec:subhead}

The inference process of DDPM is a stochastic process. So we can implicitly ensemble of segmentation masks to boost the segmentation performance. For every image of the test set, we sample 5 different segmentation masks. Then 5 different segmentation masks are ensembled by averaging and is thresholded at 0.5 to obtain a binary segmentation. In Table \ref{tab01}, the Dice score, the Jaccard index the 95 percentile Hausdorff Distance (HD95), and F1 are presented.

We conduct quantitative experiments to compare our method with a range of representative methods. Here, the U-Net \cite{10071153}, AttUnet\cite{10081650} and, U-Net++\cite{10081651} are the most representative deep learning model in the segmentation field, but none of them can estimate the uncertainty of segmentation. Bayesian U-Net\cite{10081652} and Probabilistic U-Net\cite{10241541} are representative methods that can estimate the uncertainty of segmentation. And we also compare PD-DDPM with other accelerating DDPMs, including TDPM\cite{10062040} and ES-DDPM\cite{10062041}. It should be emphasized that the size of ensemble in the comparison methods is also set to 5.

Table \ref{tab01} shows PD-DDPM (when $T'$=300) achieve the best results with respect to all four metrics. And PD-DDPM outperforming the vanilla DDPM. For visualization of the segmentation and uncertainty maps, we select three images $I_1$, $I_2$, and $I_3$ from the test set in Figure \ref{fig002}.

\begin{figure}[!t]
	\centering 
	%\vspace{-1cm}  %调整图片与上文的垂直距离
	%\setlength{\abovecaptionskip}{0.3cm}   %调整图片标题与图距离
	%\setlength{\belowcaptionskip}{0.0cm}   %调整图片标题与下文距离
	%\centerline{\includegraphics[width=0.72\textwidth]{fig03.eps}}
	%\centerline{\includegraphics[width=\columnwidth]{02.png}}
	%\centering%trim={<left> <lower> <right> <upper>}
	\includegraphics[scale=0.56,trim=0.15cm 0.4cm 0cm 0.8cm,clip]{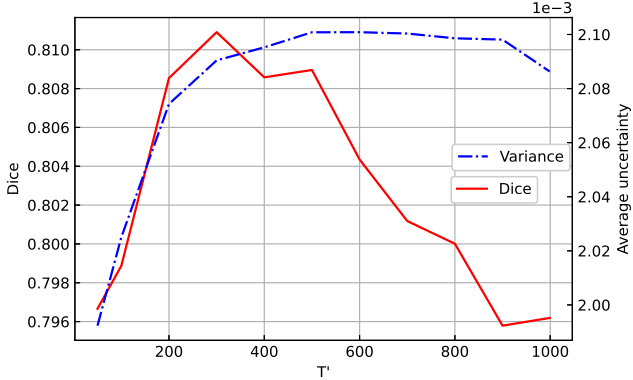}\\
	\caption{The Dice and uncertainty on testing set with respect to $T'$. The horizontal axis represents the size of $T'$.}
	\label{fig003}
\end{figure}

\subsection{Determining Optimal $T'$}
\label{ssec:subhead}

Here, we analyzed the impact of the hyperparameter $T'$ on PD-DDPM. By varying the $T'$ among \{50, 100, 200, 300, 400, 500, 600, 700, 800, 1000\}, we train PD-DDPM for WMH segmentation. As shown in Figure \ref{fig003}, PD-DDPM achieves the best Dice score when $T' = 300$. And we also analyze the effect of $T'$ on uncertainty estimation (ie, the variation of predicting softmax output). Figure \ref{fig003} shows the uncertainty increases with $T'$ when $T'$ $<$ 500. Then the $T'$ was further increased, the uncertainty tended to saturate.
 
\begin{figure}[!t]
	\centering 
	%\vspace{-1cm}  %调整图片与上文的垂直距离
	%\setlength{\abovecaptionskip}{0.3cm}   %调整图片标题与图距离
	%\setlength{\belowcaptionskip}{0.0cm}   %调整图片标题与下文距离
	%\centerline{\includegraphics[width=0.72\textwidth]{fig03.eps}}
	%\centerline{\includegraphics[width=\columnwidth]{02.png}}
	%\centering%trim={<left> <lower> <right> <upper>}
	\includegraphics[scale=0.56,trim=0.1cm 0.4cm 0cm 0.8cm,clip]{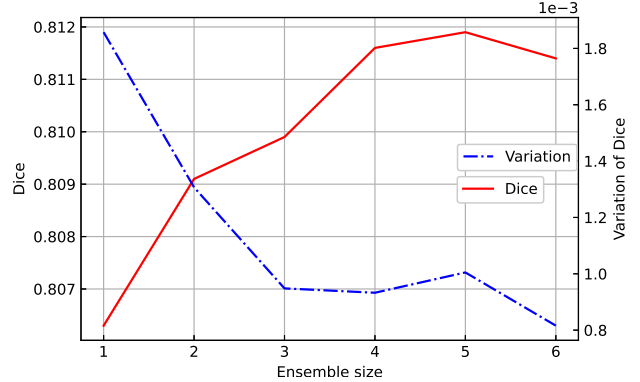}\\
	\caption{The average and standard deviation of the dice on testing set with respect to ensemble size (when $T'$=300).}
	\label{fig004}
\end{figure}

\subsection{Effect of the size of ensembles}
\label{ssec:subhead}

The optimal size of an ensemble is a task specific parameter that needs to be optimized\cite{10252041}. Figure \ref{fig004} shows that (1) the ensemble with multiple masks outperformed the only one mask. (2) when ensemble sizes increased, performance tended to saturate. We set the ensemble size to 5 in our methods. Figure \ref{fig004} also shows standard deviation of segmentation performance with respect to different ensemble sizes. The variation of segmentation performance was reduced on dice metrics when the ensemble size increased. It demonstrated that the ensemble model not only boost the segmentation performance but also guarantee a robust segmentation result.

\begin{table}
	%\footnotesize
	\centering
	\caption{Segmentation scores of PD-DDPM with different pre-segmentation Dice scores.}
	\label{table}
	\setlength{\tabcolsep}{4pt}
	\begin{tabular}{c|cccc}
		\hline 			
		Pre-segmentation accuracy       & Dice& Hd95& Jaccard& F1\\
		\hline
		%\hline
		Zero (Dice=0.000)   			&  0.353&  9.899&  0.235&  0.677\\
		%\hline
		Pre-seg (Dice=0.754)          &  0.790&  4.364&  0.662&  0.776\\
		Pre-seg (Dice=0.770)          &  0.795&  4.121&  0.667&  0.770\\
		%\hline
		Pre-seg (Dice=0.785)       	&  0.798&  3.974&  0.671&  0.786\\
		\hline
		Pre-seg (Dice=0.799)        &  \bf{0.812}&  \bf{3.494}&  \bf{0.689}&  \bf{0.800}\\
		%\hline
		\hline
	\end{tabular}
	\label{tab02}
\end{table}

\subsection{Effect of pre-segmentation accuracy}
\label{ssec:subhead}

Because the pre-segmentation result is not the real ground truth segmentation, there will be some errors between $\hat{x}^{T'}$ and $x^{T'}$. Here we analyzed the effect of pre-segmentation accuracy on PD-DDPM. As shown in Table \ref{tab02}, the higher the pre-segmentation Dice score, the higher performance of the PD-DDPM (when $T'$=300). So PD-DDPM can be combined with existing advanced segmentation networks to further improve performance and obtain uncertainty estimates.

\section{CONCLUSION}
\label{sec:typestyle}

To accelerating DDPM for medical image segmentation, this paper propose a pre-segmentation diffusion sampling DDPM (PD-DDPM), which is specially used for medical image segmentation. We empirically find that PD-DDPM achieves the best results under the parameter settings of this paper when $T'=300$ (T=1000). Experiments show even with a significantly smaller number of reverse sampling steps, PD-DDPM also outperforming the vanilla DDPM. Compared with some existing acceleration methods, our method also get the best result. Further, PD-DDPM is orthogonal to existing advanced segmentation models, which can be combined to further improve model performance and obtain uncertainty estimates.

%\bibliographystyle{IEEEbib}
%\bibliography{strings,refs}

\end{document}